# SPIRAL GALAXIES AS CHIRAL OBJECTS?

S. CAPOZZIELLO[1] and A. LATTANZI[2]

[1]*Dipartimento di Scienze Fisiche and INFN (sez. di Napoli), Università di Napoli "Federico II", Complesso Universitario di Monte S. Angelo, Via Cinthia, I-80126, Napoli, Italy*
*E-mail: capozziello@na.infn.it*
[2]*Dipartimento di Chimica, Universita' di Salerno, Via S. Allende, I- 84081, Baronissi, Salerno, Italy*

**Abstract.** Spiral galaxies show axial symmetry and an intrinsic 2D-chirality. Environmental effects can influence the chirality of originally isolated stellar systems and a progressive loss of chirality can be recognised in the Hubble sequence. We point out a preferential modality for genetic galaxies as in microscopic systems like aminoacids, sugars or neutrinos. This feature could be the remnant of a primordial symmetry breaking characterizing systems at all scales.

## 1. Introduction

Spiral galaxies can be dealt under the standard of enantiomers, i.e. objects or physical systems whose mirror images are not superimposable. This feature is easily recognized in a wide range of natural systems and it is generally reported under the standard of chiral symmetry. While it is well-known that elementary particles (e. g. L-neutrinos) or organic molecules (e. g. L-aminoacids and D-sugars) have a preferential chiral characterization, it has to be assessed if extremely large macroscopic systems as galaxies, can share this feature. Macroscopic chirality of galaxies could be related to some primordial microscopic process which led to the today observed large scale structures. In fact, a paradigm of cosmology asserts that initial quantum fluctuations have been hugely enhanced during the inflationary epoch, giving rise to the observed galaxies and clusters of galaxies, which are the largest structures in the Universe (Peacock, 1999; Liddle, 2000). The order of magnitude of such an expansion ranges from $10^{50}$ to $10^{60}$ so that, in principle, every microscopic process could have been enlarged to huge astrophysical scales (Peacock, 1999; Liddle and Lyth, 2000). On the other hand, chirality of elementary particles plays a fundamental role in physics being related to symmetry breaking which led to the violation of some fundamental conservation laws as *CP* invariance. It is therefore legitimate to ask for remnants of these processes not only in microscopic phenomena, but also in macroscopic ones, due to the cosmological expansion, with the perspective of a unitary view of physics (Barron, 1986; Avalos et al., 2002).

In this letter, we are going to develop some considerations by which spiral galaxies can be framed under the standard of molecular chirality. Some recent studies have shown that chiral tetrahedral molecules can be algebraically described by orthogonal groups (Capozziello and Lattanzi, 2003) and the approach gives evidence of common features with those of elementary particles. In particular, molecules and particles can be represented by unitary quaternions with analogous dynamics (Capozziello and Lattanzi, 2005a). Spiral galaxies are endowed with these features and a defined chiral modality could be a genetic property which galaxies are going to loose due to environmental effects.

## 2. Spirals as non-superimposable chiral objects

According to Ruch (1972), we have a genuine form of chirality, the so-called *handed* chirality (shoe-like) and the so-called *nonhanded chirality* (potato-like). In the case of tetrahedral molecules, enantiomers are handed chiral objects, while octahedral complexes, having six different ligands, can be an example of nonhanded chiral objects (King, 2001). In general, we can deal with not superimposable mirror images as chiral objects (enantiomers), neither superimposable nor mirror images chiral objects (diastereoisomers), and achiral objects (isomers which are completely superimposable by rotation). In any case, the chiral nature of objects can be established taking into account the group of transformations acting on the configuration space, i.e. the *ensemble* of all the possible configurations of the system itself. In this sense, chirality is a symmetry which emerges in abstract space (e.g. the spin-helicity configuration space of elementary particles) and it depends on the true structure of the object, rather than on the physical space where it is embedded. With this concept in mind, if a system is constituted by N elements, the group of possible transformations is $G(N)$, where $G$ is defined in the abstract space of dimension N (Capozziello and Lattanzi, 2003). A general definition of chirality is then related to the number of possible rotations and inversions which a system with N constituents can undergo. It is then straightforward to consider the $O(N)$ groups which are the groups of rotations and inversions of a given system. The case of tetrahedral molecules is enlightening and spiral galaxies, as we will see, can be dealt under the same standard.

A tetrahedron, constituted by 4 different ligands, has 4!=24 possible configurations which are nothing else but the Fischer projections of the two enantiomers of the molecule (Eliel et al., 1994).

The same enantiomer is conserved when acting on the molecule only with the elements of the subgroup *SO*(4), the 4D-special orthogonal group. The orthogonality conditions are $\frac{1}{2}4(4+1)=10$, while the number of independent generators (parameters) of the group is $\frac{1}{2}4(4-1)=6$. With this approach in mind, a chiral transformation is related to the possibility that two bonds of the molecule undergo an inversion, which, in the set of possible configurations, means a transition from the (+)-enantiomer to the (-)-enantiomer and vice-versa. From a quantum mechanical viewpoint, this can be a quantum tunneling process so that *O*(4) represents a *quantum chiral algebra* (Capozziello and Lattanzi, 2005b). It is interesting to point out that the isomorphism

$$SU(2) \times SU(2)/\pm 1 \approx SO(4) \qquad (1)$$

holds for groups *SU*(2) and *SO*(4) so that rotations and inversions of ligands can be given in terms of Pauli matrices or unitary quaternions (Capozziello and Lattanzi, 2005a). Features like these point out that quantum states of chiral tetrahedral molecules and fundamental particles show analogies which have to be seriously taken into account as previously suggested by Heisenberg (Barron, 1986). These arguments can be easily generalized to generic systems with an even number N of constituents (e.g. ligands), with N! configurations. The algebra will be *O*(N), the number of independent generators will be $\frac{N(N-1)}{2}$. Clearly the word *enantiomer* has the general meaning of a physical system, present in two modalities, with non-superimposable mirror images.

These considerations can be specified for spiral galaxies which have their arms winding in a trailing or in a leading sense (Figure 1). *Trailing* means that the tips of the arms point in the opposite direction from rotation, while in the case of *leading* arms the tips point in the same direction of rotation (Binney and Tremaine, 1987).

These features are not always immediately evident in the galaxy structure: they appear very clearly in the so-called *grand-design spirals* which usually have two symmetric and well-defined arms as in the case of the galaxy M51. However, not all spirals are grand-design with two distinct arms; they can be *flocculent* and it could be difficult to define the spiral structure.

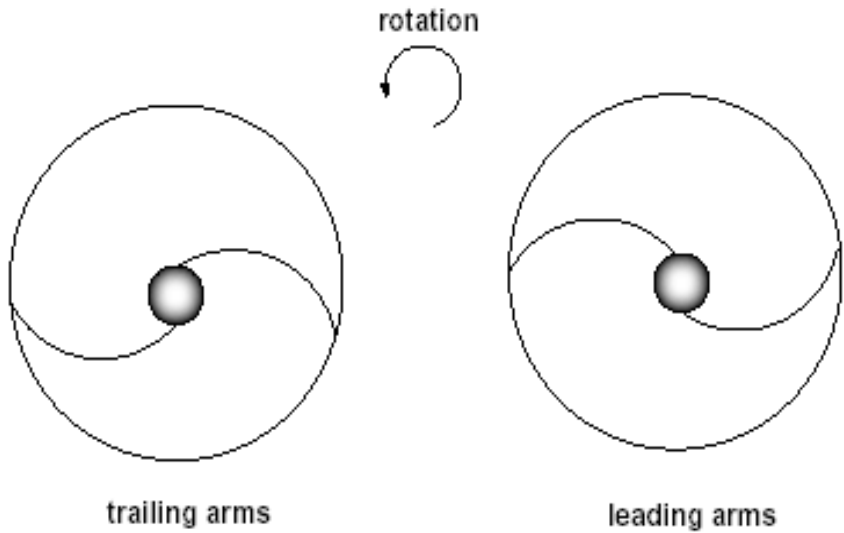

*Figure 1.* Trailing and leading modes of spiral galaxies. Spiral arms can be classified by their orientation relative to the direction of rotation of the galaxy. A trailing arm is one whose outer tip points in the direction opposite to galactic rotation, while the outer tip of a leading arm points in the direction of rotation.

The distinction between trailing and leading spiral arms requires the determination of the orientation of the plane of the galaxy relative to our line of sight so that radial velocity measurements can be unambiguously interpreted in terms of the direction of galaxy rotation. Considering the recession velocity, that is given by the Hubble relation (which holds for every galaxy thanks to the cosmological expansion) $v = H_0 d$, where $H_0$ is the Hubble constant (today esteemed ~71 Km·sec$^{-1}$·Mpc$^{-1}$) (Eidelman et al., 2004) and *d* the distance of the galaxy from the observer, a definite chirality is assigned. Moving along the arms of the galaxy toward the center and taking into account the direction of recession velocity, the helicity is assigned so that trailing galaxies are left-helical, while leading galaxies are right-helical. In almost all cases where this determination can be made, *it does appear that in general spiral arms are trailing*, (Sparke and Gallagher, 2000; Vorontsov-Vel'yaminov, 1987) while the rare leading spirals can be interpreted as the result of tidal encounters with retrograd-moving objects (the case of Andromeda galaxy M31 and its companion M32 is a remarkable example).

Turning to the orthogonal group description of chirality, the spiral arms can be considered as the ligands and the configuration space is 2D. The 2D-image (the galaxy projected onto its lying plane) cannot be superimposed with its mirror image without removing the image from the plane. The two images are the leading and trailing structures. The two modalities are parametrized by a rotation angle, then chirality is algebraically described by the O(2) group, where the number of generators is $\frac{2(2-1)}{2}=1$, corresponding to the rotation angle. To demonstrate these statements, we can write any rotation into the plane as

$$\begin{pmatrix} y' \\ x' \end{pmatrix} = \begin{pmatrix} \cos\vartheta & \sin\vartheta \\ -\sin\vartheta & \cos\vartheta \end{pmatrix} \begin{pmatrix} x \\ y \end{pmatrix} \qquad (2)$$

or, abbreviating,

$$x^{i'} = O^{ij}(\vartheta) x^j \qquad (3)$$

with $x^1 = x$ and $x^2 = y$ where $O^{ij}(\vartheta)$ is the rotation matrix. Being the group orthogonal, we have $O^{-1}(\vartheta) = O(-\vartheta)$ and then

$$O^{-1}(\vartheta)O(-\vartheta) = \mathbf{I} = \begin{pmatrix} 1 & 0 \\ 0 & 1 \end{pmatrix} \qquad (4)$$

which means that the inverse of an orthogonal matrix is its transpose. Any orthogonal matrix can be written as the exponential of a single antisymmetric matrix $\tau$, that is

$$O(\vartheta) = e^{\vartheta \cdot \tau} \equiv \sum_{n=0}^{\infty} \frac{1}{n!}(\vartheta\tau)^n \qquad (5)$$

where

$$\tau = \begin{pmatrix} 0 & 1 \\ -1 & 0 \end{pmatrix} \qquad (6)$$

so then we have that the transpose of $e^{\vartheta \cdot \tau}$ is $e^{-\vartheta \cdot \tau}$:

$$O^T = (e^{\vartheta \cdot \tau})^T = e^{-\vartheta \cdot \tau} = O^{-1} \qquad (7)$$

Another way to prove this identity is to power expand the right-hand side and sum up the series. We obtain

$$e^{\vartheta \cdot \tau} = \cos\vartheta \cdot \mathbf{I} + \tau\sin\vartheta = \begin{pmatrix} \cos\vartheta & \sin\vartheta \\ -\sin\vartheta & \cos\vartheta \end{pmatrix} \qquad (8)$$

which demonstrates that all elements of O(2) are parametrized by one angle $\vartheta$, so we can say that O(2) is a one-parameter group.

Let us now take the determinant of both sides of the defining equation (7):

$$\det(OO^T) = \det O \det O^T = (\det O)^2 = 1 \qquad (9)$$

which means that the determinant of *O* is equal to ±1. If we take $\det(O) = 1$, the resulting subgroup is *SO*(2), the special orthogonal matrices in two dimensions describing 2D-rotations. Another subset of *O*(2) is that defined by $\det(O) = -1$. It consists of elements of *SO*(2) times the matrix

$$\begin{pmatrix} 1 & 0 \\ 0 & -1 \end{pmatrix} \quad (10)$$

corresponding to discrete transformations as

$$\begin{aligned} x &\to x \\ y &\to -y \end{aligned} \quad (11)$$

which take a plane and map it into its mirror image. For galaxies, the two *enantiomers* are *mirror* trailing and leading structures and, in order to pass from one to the other, a discrete inversion is necessary (Figure 1). In this sense, spiral galaxies are *non-superimposable chiral objects*.

Another important remark has to be done at this point. Let us take a complex number $u = a + ib$. It can be transformed as

$$u' = U(\vartheta)u = e^{i\vartheta}u \quad (12)$$

where $U(\vartheta)$ is a complex unitary matrix

$$U \times U^\dagger = \mathbf{I} \quad (13)$$

The set of all one-dimensional unitary matrices $U(\vartheta) = e^{i\vartheta}$ is the group *U*(1) where the multiplication law

$$e^{i\vartheta}e^{i\vartheta'} = e^{i\vartheta + i\vartheta'} \quad (14)$$

holds. This multiplication law is the same of *O*(2) even though this construction is based on a new space of complex one-dimensional numbers. Then we have the correspondences

$$SO(2) \approx U(1), \qquad e^{\vartheta \cdot \tau} \leftrightarrow e^{i\vartheta} \quad (15)$$

which means that two real numbers which transform under *O*(2) can be combined into a single complex number (or function) which transforms under *U*(1).

As it is well known, the *U*(1) group describes the transformations of the electromagnetic field. In the present context, it describes the symmetry between trailing and leading modes which transform each other by complex conjugation. In fact, if $U(\vartheta) = e^{i\vartheta}$ describes a trailing mode, the complex conjugate one, $U^\dagger(-\vartheta) = e^{-i\vartheta}$, represents a leading mode.

### 3. Discussion

Summarizing these results, spiral galaxies (at least the grand-design ones) are enantiomers which *naturally* appear in the left-helical series characterized by trailing arms. These structures are *quasistatic density waves* which are stable with respect to the typical age of a galaxy (~$10^{10}$ yr) and take part to the global motion around the bulge with periods of the order of $10^8$ yr. These results constitute the solution, previously obtained by Lin and Shu (Lin and Shu, 1964, 1966), of the so-called *winding problem* which led to the paradox that *material arms* composed by a fixed set of identifiable stars and gas clouds would necessarily *wind up* on a short time scale if compared to the age of the galaxy. Considering density waves with a *global pattern speed* (Elmegreen, 1998; Binney and Tremaine, 1987; Binney and Merrifield 1998), the spiral structure results stable against small perturbations and remains permanent during the evolution unless the galaxy would undergo strong gravitational interactions. This feature of a spiral galaxy can be considered *genetic* in the sense that it is the original angular momentum of the protogalaxy that gives rise to the spiral structure which permanently remains a characteristic of the system. Joined to the fact that, observations almost always reveal trailing arms, (Sugai and Iye, 1991; Hu, 1998; Aryal and Saurer, 2004, 2005; Kashikawa and Okamura, 1992) it seems that *chiral asymmetry* is present in spiral galaxies and only one modality is favoured likewise in the case of aminoacids and neutrinos (both left-handed).

This statement deserves a wide discussion, first of all in connection to the other morphological types of Hubble sequence and the environmental effects which could lead toward a progressive loss of *chirality* in galaxies.

Recently, Kondepudi and Durand, (2001) performed a statistical analysis taking into account the spiral galaxies in the Carnegie Atlas of Galaxies (Sandage and Bedke, 1996). They considered 540 galaxies classified as normal (**S**) or barred (**SB**) spirals. An interesting dominance of trailing structures emerges for **S**-type galaxies and a dominance of leading structures is revealed for **SB**-type galaxies. In other words, the presence of the bar is mostly correlated to the leading modes, while grand-design normal spiral galaxies are trailing. This fact can be interpreted considering the relative positions of morphological Hubble types in the clusters of galaxies and the probability that galaxies suffer strong or tidal gravitational interactions among them. A wide number of studies (Busarello et al., 1997; Sparke and Gallagher, 2000; Vorontsov-Vel'yaminov, 1987) report that elliptical (**E**) and lenticular (**S0**) galaxies reside in large clusters, typically containing from $10^2$ to $10^3$ galaxies, while spiral galaxies are found on the boundary of clusters, in loose groups or in open fields. Clusters are often dense environments, bound together by the mutual gravitational attraction of constituent galaxies which is going to increase toward the center. Following Bautz and Morgan (1970), clusters of galaxies can be classified considering the relative contrast of the brightest member galaxy. Symmetric, well-shaped clusters are core-dominated by huge elliptical **cD** galaxies (with masses ranging from $10^{13}$ to $10^{14}$ solar masses) which site in the center of symmetry (e.g. Figure 2). These clusters are classified as BM type I. Clusters where brightest galaxy (or galaxies) are intermediate in appearance between **cD** and the Virgo-type giant ellipticals are BM type II. Finally, clusters containing no dominant galaxies are BM type III. However, intermediate BM type I-II and BM type II-III have to be taken into account since the classification is not so sharp. In general, for BM types I, I-II, and II clusters, spiral galaxies reside in less dense zones, where mutual gravitational interactions are very rare. BM types II-III and III can be further subdivided: for example, BM type III-**E** and BM type III-**S** mean respectively the absence or the presence of considerable numbers of bright spirals. **E** and **S0** galaxies tend to reside in high density regions (i.e., near the cluster plane in general sense) of the clusters, if the large-scale morphology segregation is present (Dressler, 1980; Giovanelli et al., 1986). On the other hand, **S** galaxies tend to reside outside the cluster plane, toward the edges. The **SB** galaxies are in an intermediate situation. Bars are the result of gravitational interactions not so strong to destroy the global spiral structure. As shown by numerical simulations by Elmegreen et al. (1990) and Gerin et al. (1990) the interaction of two galaxies gives rise to tidal tensions capable of twisting the buldge and inducing the formation of a bar. Unlike spiral arms, bars are not density waves since the majority of their stars always remains within the bar. This means that, although photometric observations suggest that bars and spiral arms seem joined, the former are leading structures in counter-rotation with respect to the spiral arms. The cumulative effect, as in the case of M31 and M32, would be that a leading spiral structure results and the morphological type evolves as **S** $\Rightarrow$ **SB**. Without entering into details of bar formation, **SB** galaxies can be seen as an intermediate state between chirally-defined systems (**S**-galaxies) and system which are loosing, by gravitational interactions, their chiral features (**E** and **S0** galaxies) since they are living in much dense environments. Gravitational field and tidal encounters act as a sort of thermic or photo-induced racemization for enantiopure compounds (Rayner et al., 1968; Mislow et al., 1965). In other words, starting from a well-defined chiral modality (trailing **S**-galaxies), an evolution through a pseudo-racemic mixture (**S** and **SB**-galaxies) and toward completely achiral objects (**E**, **S0** galaxies) can be envisaged. **E** and **S0** galaxies can be considered isomers which are completely superimposable by rotation. An important remark is due at this point. It has to be noted that the **S** and **SB** galaxies are supported by rotation while **E** galaxies are supported by stellar random velocity dispersion, rather than by rotation (Kashikawa and Okamura, 1992). This means that the progressive loss of chirality could be related to the fact that rotationally-supported stellar systems evolve toward randomised stellar systems thanks to dynamical processes, as merger, which globally give rise to a *dynamical loss of chirality*. A pictorial scheme of the situation is sketched in Figure 2, where the *genetic chirality* is going to be lost when the systems are evolving toward high density zones in a BM type I or II cluster of galaxies. Another interesting result is reported by Aryal and Saurer (2005). These authors analysed the distribution of angular momentum of 2280 **S** galaxies and 1162 **SB** galaxies in the Local

Supercluster. They found anisotropy for spirals and no preferred orientation for **SB.** For the subsamples of late types **S** and **SB,** they found anisotropy. Also this fact indicates the possibility of dominance of trailing or leading structures.

## 4. Conclusions

In conclusion, it appears that the Hubble morphological sequence is related to the degree of chirality of galaxies and, we can say that late-type galaxies (**S**) are chiral objects, while early-type galaxies (**E**) are achiral objects. In this perspective, the Hubble diagram is a chirality loss sequence indicating the dynamical evolution of the stellar systems.

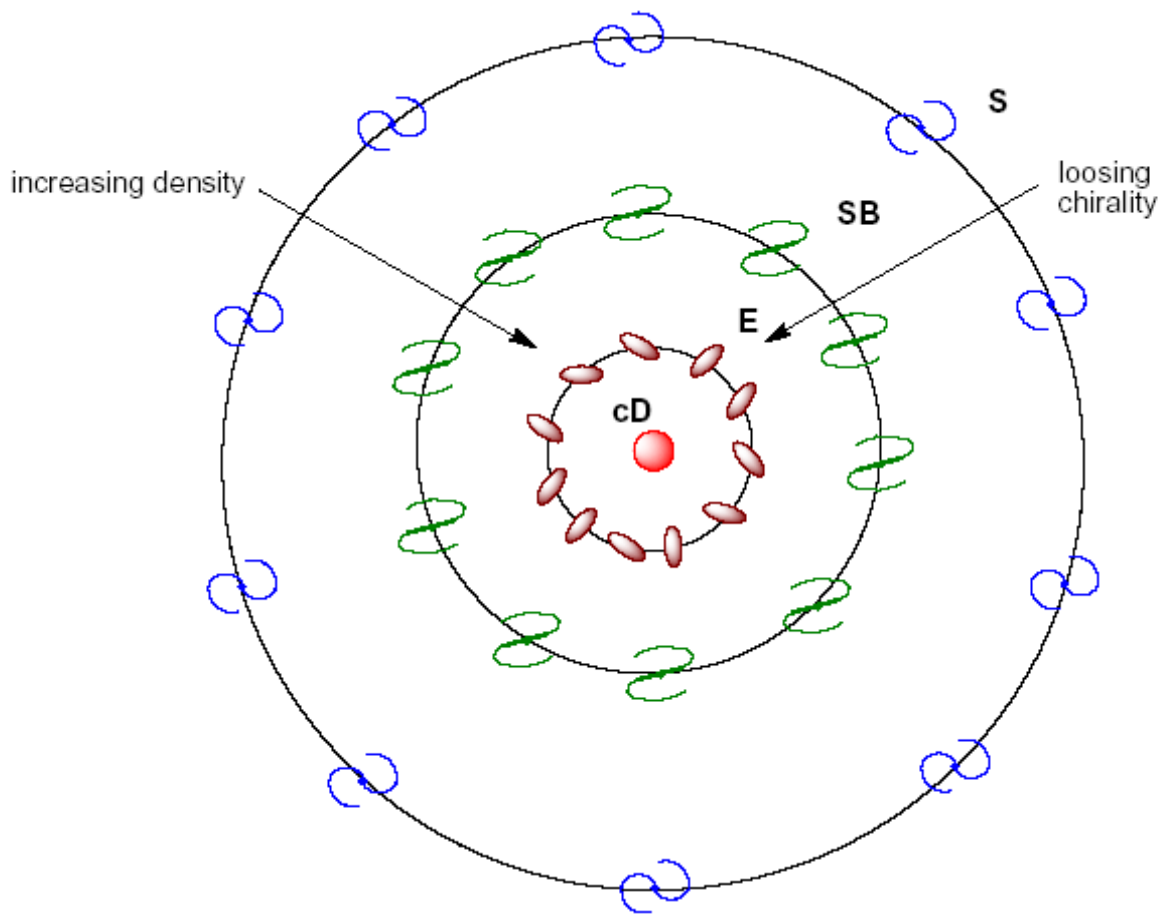

*Figure 2*. Pictorial view of morphology-density relation indicating the loss of chirality toward the center of a cluster of galaxies. This sketch could realistically represent the situation in the cases of types I and II-clusters of the Bautz and Morgan classification, where the large-scale morphology segregation and the (**cD** or Virgo-type) central giant galaxies are present. In the case of type-III clusters, a dynamical loss of chirality is difficult to be envisaged

However, our hypothesis on a preferred genetic modality of chiral galaxies has to be more confidentially supported by deeper and refined analysis of large observational data sets (e.g. 2dF or SDSS surveys), as soon as their morphological characters will be better defined.